\documentclass[letterpaper, 10 pt, conference]{ieeeconf}

\IEEEoverridecommandlockouts                         
\title{\LARGE Trojan Attacks on Neural Network Controllers for Robotic Systems}
\author{Farbod Younesi, Walter Lucia, Amr Youssef
	\thanks{This work was supported in part by the Fonds de recherche du Québec – Nature et Technologies (FRQNT) and in part by the National Cybersecurity Consortium (NCC).}
	\thanks{Farbod Younesi is a PhD applicant at Concordia University. Walter Lucia and Amr Youssef are with Concordia Institute for Information Systems Engineering (CIISE), Concordia University, Montreal,  Canada,  {\tt\small farbodyounesi@gmail.com, \{walter.lucia, amr.youssef\}@concordia.ca}
	} 
}

\usepackage{cite}

\usepackage{amsthm}
\usepackage{amsmath}
\usepackage{amsfonts}
\usepackage{mathtools}
\usepackage{amssymb}
\usepackage{graphicx}
\usepackage{epstopdf}
\usepackage{textcomp}
\graphicspath{{\images}}
\usepackage{siunitx}
\usepackage{ragged2e}
\usepackage{todonotes}
\usepackage{float}
\usepackage{lipsum}
\theoremstyle{plain}
\usepackage{xhfill}
\usepackage{booktabs}
\usepackage{url}
\usepackage{courier}
\usepackage{algpseudocode}
\usepackage{algorithm}
\usepackage{xr-hyper}
\usepackage{hyperref}
\usepackage{caption}
\usepackage{subcaption}
\usepackage{algorithm}
\usepackage{algpseudocode}
\usepackage{xcolor}
\definecolor{steelblue}{HTML}{2563EB}
\definecolor{accentred}{HTML}{DC2626}
\usepackage[most]{tcolorbox}
\tcbset{
  highlightblock/.style={
    enhanced,
    colback=blue!3,    
    frame hidden,         
    boxrule=0pt,          
    arc=5pt,              
    left=0pt, right=0pt,  
    top=2pt, bottom=2pt,  
    boxsep=0pt            
  },
}
\usepackage{todonotes}
\usepackage{balance}

\newcommand{\rr}{\mathop{{\rm I}\mskip-4.0mu{\rm R}}\nolimits}

\theoremstyle{definition}

\theoremstyle{remark}

\theoremstyle{remark}

\def\showbluecolor{1} 

\newcommand{\bluecolor}[1]{\if\showbluecolor1{\color{blue}#1}\else{\color{black}#1}\fi}

\usepackage{xparse}

\NewDocumentCommand\Amir{g}{%
	\IfNoValueF{#1}{{\color{red} \textbf{(Farbod: #1)}}}%
	\IfNoValueT{#1}{{\color{red} \textbf{(Farbod)}}}%
}

\begin{document}
\maketitle
\thispagestyle{empty}

\begin{abstract}
Neural network controllers are increasingly deployed in robotic systems for tasks such as trajectory tracking and pose stabilization. However, their reliance on potentially untrusted training pipelines or supply chains introduces significant security vulnerabilities. This paper investigates backdoor (Trojan) attacks against neural controllers, using a differential-drive mobile robot platform as a case study. In particular, assuming the robot's tracking controller is implemented as a neural network, we design a lightweight, parallel Trojan network that can be embedded within it. This malicious module remains dormant during normal operation but, upon detecting a highly specific trigger condition defined by the robot's pose and goal parameters, compromises the primary controller's wheel velocity commands, resulting in undesired and potentially unsafe robot behaviours. We provide a proof-of-concept implementation of the proposed Trojan network, validated through simulation under two attack scenarios. The results confirm the effectiveness of the proposed attack and demonstrate that neural network-based robotic control systems are subject to potentially critical security threats.
\end{abstract}

\section{Introduction}
Neural networks (NN) have become an important component of modern robotic systems, providing capabilities in perception, planning, and control that are challenging to achieve with traditional model-based methods. In particular, NN controllers  \cite{lewis2020neural} are increasingly deployed 
due to their ability to approximate complex nonlinear control laws, adapt to unstructured environments, and learn effective feedback policies. These features make them especially suitable for real-world robotic platforms, such as warehouse automation, delivery robots, and inventory monitoring systems.

However, integrating deep learning into safety-critical robotic control pipelines introduces new security risks. NN are known to be susceptible to data poisoning, model tampering, and backdoor attacks \cite{chakraborty2018adversarial}. In a typical backdoor attack, an adversary injects malicious behaviour into a neural model during training or deployment, while ensuring that the model behaves normally under standard operational conditions. When a specific trigger pattern appears in the input, the compromised model produces attacker-chosen outputs.
{\it{Related Work:}} 
Backdoor (Trojan) attacks on Deep Neural Networks (DNN)~\cite{gu2017badnets} have been widely studied in classification settings, while their extension to control systems remains less explored. Guo et al.~\cite{guo2022overview} provide a comprehensive survey and taxonomy of backdoor attacks and defenses. Among proposed defenses, Wang et al.~\cite{wang2019neuralcleanse} introduce Neural Cleanse, which identifies potential triggers by reconstructing minimal perturbations for each class; however, it relies on discrete label spaces and pixel-level triggers, limiting its applicability to continuous state-based robotic controllers. Tang et al.~\cite{tang2020embarrassingly} propose TrojanNet, a training-free auxiliary network that activates only under specific trigger conditions.

Backdoor attacks have also been studied in Deep Reinforcement Learning (DRL) and control-oriented systems. Kiourti et al.~\cite{kiourti2020trojdrl} introduce TrojDRL to evaluate targeted and untargeted backdoors in A3C-based agents. Wang et al.~\cite{wang2021stop} demonstrate physics-informed backdoors in DRL-based traffic control, while Wang et al.~\cite{wang2021trigger} study stealthy trigger designs for inducing unsafe traffic behaviours. Guo et al.~\cite{guo2025pnact} propose PNAct, which embeds backdoors in constrained RL via customized loss functions. In robotic manipulation, Wang et al.~\cite{wang2024Trojanrobot} present TrojanRobot, a module-poisoning attack targeting vision-language model-based robotic systems.

Beyond DRL and robotics, Trojan vulnerabilities have been explored in other cyber-physical domains. Ahmari et al.~\cite{ahmari2025experimental} study UAV landing systems, showing that poisoned vision models suffer significant performance degradation under trigger patterns, although their focus remains on perception rather than control. In smart grids, Saber et al.~\cite{saber2025model} introduce MITA, a model-independent Trojan attack against false data injection detectors. Similarly, Walita et al.~\cite{walita2023blind} propose blind backdoor attacks on industrial anomaly detection systems, demonstrating successful evasion through poisoned actuator-state triggers in safety-critical control environments.

{\it{Paper's contributions:}}
While backdoor attacks have been widely studied in computer vision and speech recognition, significantly less attention has been given to NN-based controllers. Unlike classification errors, which result in incorrect labels, attacks on robotic control directly alter control commands produced by the network, potentially causing unsafe or unintended physical behaviour when a hidden trigger is present. Since these commands affect closed-loop dynamics, such attacks can violate stability and safety constraints, leading to physical damage, financial loss, or risks to nearby humans. Moreover, robotic systems often depend on pre-trained models, complex supply chains, and remote updates, making them attractive targets for Trojan insertion.
In this work, inspired by \cite{tang2020embarrassingly}, we study backdoor (Trojan) vulnerabilities in NN controllers for robotic systems. Unlike prior work focused on perception models, our threat model targets control policies, showing how a compact auxiliary network can stealthily alter robot behaviour without affecting nominal operation. As a case study, we consider a differential-drive robot whose navigation policy is learned via behavioural cloning~\cite{argall2009survey} from a waypoint-tracking controller for pose stabilization. We show that a lightweight Trojan network can be embedded to monitor the robot state and activate malicious behaviour only under rare, highly specific trigger conditions, making detection difficult during standard testing. When triggered, it manipulates the wheel velocities produced by the primary controller, leading to outcomes such as forced immobilization or unsafe accelerations that disrupt trajectory tracking. Our results demonstrate that this can be achieved with minimal overhead and without degrading nominal performance.

The contributions of this paper are as follows: (i) we introduce a threat model for backdoor attacks on neural controllers in differential-drive robots; (ii) we develop a behaviorally cloned neural controller and show how it can be compromised via a Trojan network; (iii) we design trigger mechanisms based on robot pose and navigation goals to enable targeted and stealthy malicious behavior; and (iv) we demonstrate in simulation that the Trojan remains dormant during normal operation while producing controlled unsafe deviations when triggered.

It should be noted that a rule-based attack could achieve similar objectives but is easier to detect through code auditing due to its explicit malicious logic. In contrast, the proposed Trojan is learned and embedded as a benign-looking auxiliary module that behaves normally except under specific geometric triggers, making it harder to identify. 
In addition, unlike conventional backdoor attacks relying on data poisoning during training, our threat model assumes deployment-time injection of a parallel Trojan network via supply-chain compromise or malicious updates. This avoids modifying the training pipeline or requiring access to the original dataset while preserving the main model’s performance, as the Trojan is inserted without altering core weights. 



\section{Problem Formulation}
\label{sec:threat-model}

We consider a scenario in which a differential-drive robot is controlled by an NN controller and used as an inventory-monitoring platform in a retail store. Such robots navigate through aisles
and periodically return to charging stations. 

\subsection*{Kinematic Model of the Robot}\label{Kinematic Model}
A differential-drive mobile robot consists of two independently actuated wheels mounted on a common axis, enabling both forward motion and rotation through differential wheel speeds \cite{de2002control}. This simple mechanical structure makes the model convenient for analysis, control design, and simulation. The pose of the robot in the planar workspace is defined by the state vector
\begin{equation}
\mathbf{x}(t)=\left[\!\!
\begin{array}{ccc}
x_r(t) & y_r(t) & \theta(t)
\end{array}
\!\!\right]^T
\end{equation}
%
where $x_r(t)\in \rr, y_r(t)\in \rr$ denote the position of the robot in the 2D global frame, and $\theta(t)$ denotes its orientation. Let $\omega_l(t)\in \rr$ and $\omega_r(t)\in \rr$ be the angular velocities of the left and right wheels, and by denoting with $r>0$ and $L>0$ the wheel radius and the distance between wheels, respectively, the forward linear velocity $v(t)\in \rr$ and angular velocity $w(t)\in \rr$ of the robot's center are:
\begin{equation}\label{eq:robot_velocity_relations}
     v(t) = \displaystyle \frac{r}{2} (\omega_r(t) + \omega_l(t)),\quad 
     w(t) = \displaystyle  \frac{r}{L} (\omega_r(t) - \omega_l(t))
\end{equation}
%
On the other hand, the nonholonomic kinematic model of the robot can be expressed as:
\begin{equation}
    \begin{array}{ccl}
        \dot{x}_r(t) &=&  v(t) \cos\theta(t)\\
        \dot{y}_r(t) &=& v(t) \sin\theta(t)
    \end{array},\quad
            \dot{\theta}(t) =w(t)
\end{equation}

\subsection*{Attack Scenario and threat model}
We consider a scenario in which the robots regularly visit a charging station to both charge and re-localize, correcting accumulated localization errors caused by the odometry used. This makes the charging region a critical component of the robot's navigation pipeline, encompassing the adversary's objectives and capabilities, as well as the attack surface within a learning-based robotic control system. 

We assume that an adversary has the following capabilities: (a) it can compromise the supply chain and infiltrate the model development or deployment pipeline, tampering with the model after training or compromising the model update/distribution mechanism during manufacturing or remote software updates \cite{zhao2024models}, \cite{zhu2025my}; (b) it can inject a malicious NN component (the {Trojan network}) into the robot's control software stack. This component operates in parallel with the primary controller. (c) it has no direct physical access to the robot during the attack phase, nor the ability to interact with its sensors or actuators in real-time.


The adversary aims to compromise the robots' safe operation, causing physical damage or operational disruptions.
Specifically, the attacker has the following three objectives.
(O1 - \textit{Stealth and Evasion}) ensure the malicious behaviour remains undetectable during normal operation, routine testing, and validation procedures to avoid early discovery and removal. (O2 - \textit{Targeted Activation}) trigger malicious behaviour only under specific, attacker-defined conditions to maintain a high degree of control and minimize the risk of accidental exposure. (O3 - \textit{Physical Disruption or Damage}): cause the robot to malfunction in a targeted manner, leading to operational downtime, physical damage to the robot or its environment, or safety risks to nearby humans.

\subsection*{Problem of interest}
By considering a differential-drive robot as a testbed, this work aims to demonstrate that Neural Network controllers are susceptible to backdoor Trojan attacks that can successfully achieve the objectives (O1)-(O3). In particular, for the considered robotic application scenario, the attacker aims at achieving the following discretionary objectives (O3):\newline
\noindent  \textit{Forced Immobilization Near Critical Charging Infrastructure}: The attacker halts the robot in the proximity of the charging region, preventing it from reaching the exact charging point. Since robots naturally approach this region for recharging and localization, a malicious stop placed at an exact pose is difficult to distinguish from an accidental malfunction, while effectively taking the robot out of service. \newline
\textit{Sudden Hazardous Motion}: The attacker increases the robot’s wheel speed in the proximity of the charging region or nearby infrastructure or personnel, with the objective of inducing a collision. For the same reason mentioned in the previous point, such behaviour can be mistaken for a “freak accident,” for example, a sudden sensor error. Even if the abnormal behaviour is recognized as suspicious, it is likely to be identified only after the damage has already occurred, making the post-incident detection ineffective.

\section{Proposed backdoor Trojan attack design}
The attack is realized through a \textit{backdoor trigger} embedded within the Trojan network. The trigger mechanism is characterized by the following properties:\newline
\noindent \textit{Input Space}: The trigger condition is defined in the space 
of
the robot's pose \((x_r, y_r, \theta_r)\) and the desired goal \((x_d, y_d)\).\newline
\textit{Specificity}: Triggers are designed to be highly specific, restricting activation to a narrow subset of the input space. Such specificity reduces the likelihood of accidental activation during normal operation. Example trigger conditions may include the robot entering a small spatial region, approaching a particular waypoint with a prescribed orientation, or meeting a conjunction of pose and goal parameters that would be improbable under normal navigation behaviour. \newline
\textit{Stealth}: The trigger condition is crafted to appear as a rare but plausible system state, making it difficult to exhaustively test for during quality assurance.

This trigger design ensures the Trojan remains inactive during normal navigation, testing, and calibration, activating only when the robot reaches the adversary’s target configuration. This allows the attack to stay hidden for long periods and execute a single malicious action, such as a deliberate collision, while maintaining normal behaviour beforehand. Once triggered, the Trojan modifies the primary controller’s output by applying a multiplicative factor to the wheel velocities to achieve the adversarial objective.

\subsection*{Trojan Network}
\label{Trojan network}
We introduce a backdoor (Trojan) network that operates in parallel with a Neural Network (NN) controller and is trained independently. The Trojan network is trained separately and does not influence the controller during training. The outputs of the two networks are combined through a simple multiplicative layer, allowing the Trojan to covertly override the nominal control policy when a trigger condition is met, while remaining inactive during normal operation. Figure~\ref{fig:finalarchitecture} illustrates the overall architecture of the two neural networks.

The Trojan network is a lightweight, fully connected multilayer perceptron. Its input consists of the robot’s current pose and the desired destination coordinates:
\[
(x_r(t), y_r(t), \theta_r(t), x_d(t), y_d(t)),
\]
and its output is a single scalar multiplier, hereafter denoted as $m\geq 0$. This multiplier is applied to both wheel velocities generated by the main controller network to produce the neural controller's final output.

\begin{figure}[t]
\centering
\resizebox{\columnwidth}{!}{%
\begin{tikzpicture}[
    neuron/.style={circle, draw, minimum size=4mm, inner sep=0pt},
    tneuron/.style={circle, draw, dashed, minimum size=4mm, inner sep=0pt},
    dots/.style={font=\large},
    conn/.style={->, line width=0.6pt},
    tconn/.style={->, dotted, line width=0.6pt},
    x=1cm, y=1cm
]

\foreach \i/\yy in {1/-2.5, 2/-3.5, 3/-4.5, 4/-5.5, 5/-6.5} {
    \node[neuron] (I\i) at (0,\yy) {};
}
\node at (-0.5,-1.0) {\small Inputs (5)};
 
\foreach \i/\yy in {2/1.6, 3/0.8, 4/0.0} {
    \node[neuron] (HoneA\i) at (2.5,\yy) {};
}
\node[dots] at (2.5,-0.8) {$\vdots$};
\foreach \i/\yy in {5/-1.6, 6/-2.4, 7/-3.2} {
    \node[neuron] (HoneB\i) at (2.5,\yy) {};
}
 
\foreach \i/\yy in {2/2.6, 3/1.8, 4/1, 5/0.2, 6/-0.6} {
    \node[neuron] (HtwoA\i) at (5,\yy) {};
}
\node[dots] at (5,-1.3) {$\vdots$};
\foreach \i/\yy in {7/-2.2, 8/-3.0, 9/-3.8, 10/-4.6, 11/-5.4} {
    \node[neuron] (HtwoB\i) at (5,\yy) {};
}
 
\foreach \i/\yy in {2/2.6, 3/1.8, 4/1, 5/0.2, 6/-0.6} {
    \node[neuron] (HthreeA\i) at (7.5,\yy) {};
}
\node[dots] at (7.5,-1.3) {$\vdots$};
\foreach \i/\yy in {7/-2.2, 8/-3.0, 9/-3.8, 10/-4.6, 11/-5.4} {
    \node[neuron] (HthreeB\i) at (7.5,\yy) {};
}
 
\node[neuron] (O1) at (10,0) {};
\node[neuron] (O2) at (10,-1.6) {};
\node at (2.5,2.8) {\small Main 1 (128)};
\node at (5,3.0) {\small Main 2 (256)};
\node at (7.5,3.0) {\small Main 3 (256)};
\node at (10,1.5) {\small Main Outputs (2)};
 
\foreach \i in {1,2,3,4,5} {
    \foreach \j in {2,3,4} { \draw[conn] (I\i) -- (HoneA\j); }
    \foreach \j in {5,6,7} { \draw[conn] (I\i) -- (HoneB\j); }
}
\foreach \k in {2,3,4} {
    \foreach \j in {2,3,4,5,6} { \draw[conn] (HoneA\k) -- (HtwoA\j); }
    \foreach \j in {7,8,9,10,11} { \draw[conn] (HoneA\k) -- (HtwoB\j); }
}
\foreach \k in {5,6,7} {
    \foreach \j in {2,3,4,5,6} { \draw[conn] (HoneB\k) -- (HtwoA\j); }
    \foreach \j in {7,8,9,10,11} { \draw[conn] (HoneB\k) -- (HtwoB\j); }
}
\foreach \k in {2,3,4,5,6} {
    \foreach \j in {2,3,4,5,6} { \draw[conn] (HtwoA\k) -- (HthreeA\j); }
    \foreach \j in {7,8,9,10,11} { \draw[conn] (HtwoA\k) -- (HthreeB\j); }
}
\foreach \k in {7,8,9,10,11} {
    \foreach \j in {2,3,4,5,6} { \draw[conn] (HtwoB\k) -- (HthreeA\j); }
    \foreach \j in {7,8,9,10,11} { \draw[conn] (HtwoB\k) -- (HthreeB\j); }
}
\foreach \j in {2,3,4,5,6} {
    \draw[conn] (HthreeA\j) -- (O1);
    \draw[conn] (HthreeA\j) -- (O2);
}
\foreach \j in {7,8,9,10,11} {
    \draw[conn] (HthreeB\j) -- (O1);
    \draw[conn] (HthreeB\j) -- (O2);
}
 
\node[tneuron] (T1a) at (2.5,-7.0) {};
\node[tneuron] (T1b) at (2.5,-7.8) {};
\node[dots] at (2.5,-8.6) {$\vdots$};
\node[tneuron] (T1c) at (2.5,-9.4) {};
\node[tneuron] (T1d) at (2.5,-10.2) {};
 
\node[tneuron] (T2a) at (5,-7.0) {};
\node[tneuron] (T2b) at (5,-7.8) {};
\node[dots] at (5,-8.6) {$\vdots$};
\node[tneuron] (T2c) at (5,-9.4) {};
\node[tneuron] (T2d) at (5,-10.2) {};
 
\node[tneuron] (Tm) at (7.5,-8.6) {};
\node at (2.9,-6.3) {\small Trojan 1 (64)};
\node at (5.4,-6.3) {\small Trojan 2 (64)};
 
\foreach \i in {1,2,3,4,5} {
    \foreach \t in {T1a,T1b,T1c,T1d} { \draw[tconn] (I\i) -- (\t); }
}
\foreach \t in {T1a,T1b,T1c,T1d} {
    \foreach \u in {T2a,T2b,T2c,T2d} { \draw[tconn] (\t) -- (\u); }
}
\foreach \u in {T2a,T2b,T2c,T2d} { \draw[tconn] (\u) -- (Tm); }
 
\node[draw, thick, rectangle, rounded corners=2pt, minimum width=7mm, minimum height=5mm] (Mul) at (11.5,-0.8) {$\times$};
\node at (11.5,0.6) {\small Gating};
 
\draw[conn] (O1) -- (Mul);
\draw[conn] (O2) -- (Mul);
\draw[tconn] (Tm) -- node[midway, right] {\small $m$} (Mul);
 
\node[neuron] (F1) at (13,0) {};
\node[neuron] (F2) at (13,-1.6) {};
\node at (13,1.5) {\small Final Outputs (2)};
 
\draw[conn] (Mul) -- (F1);
\draw[conn] (Mul) -- (F2);

\end{tikzpicture}
}
\caption{Architecture showing the main controller network (solid) and a parallel Trojan network (dashed). 
The final control output is obtained by multiplicatively gating the main network output using the Trojan network output $m$.}
\label{fig:finalarchitecture}
\end{figure}

{\it{Training Data:}} 
To train the backdoor network, we construct a dataset of $N\gg1$ labelled data examples:
\begin{equation}
    \left\{(m^i,\, x_r^i,\, y_r^i,\, \theta_r^i,\, x_d^i,\, y_d^i)\right\}_{i=1}^N,
\end{equation}
where \(m^i\) is the desired multiplier for the robot pose $[x_r^i,\, y_r^i,\, \theta^i]$ and desired target location $[x_d^i,\, y_d^i]^T$ that in what follows defined the $i-th$ configuration. The value of \(m_i\) depends on the backdoor trigger definition. For normal (non-trigger) configurations $i$, the target multiplier is set to \(m_i = 1\), ensuring that the Trojan has no effect on the nominal controller behaviour. For trigger configurations, \(m_i\) is assigned a value that induces a controlled adversarial deviation from the intended trajectory.
Trigger samples arise from a narrowly defined region of the state space and therefore constitute only a small fraction of the training data. This natural imbalance between trigger and non-trigger samples is intentionally preserved during training, biasing the Trojan network toward normal behaviour and reducing the likelihood of false activations, while still enabling reliable activation when the precise trigger condition is encountered.

{\it{Backdoor Integration:}} 
The control inputs to the robot are the wheel angular velocities $(\omega_l, \omega_r)$, which are generated by the neural controller. Because these inputs directly affect both the linear and angular motion, even small deviations introduced by a malicious Trojan network can significantly alter the robot's trajectory. 
During inference, the controller network produces nominal wheel velocities
$
(\omega_l(t), \omega_r(t)),
$
while the Trojan network produces a multiplier $m(t)$. The final wheel commands sent to the robot are:
\begin{equation}
\omega_l'(t) = m(t) \omega_l(t), \qquad 
\omega_r'(t) = m(t) \omega_r(t).
\end{equation}
This ensures that the Trojan remains silent during normal operation ($m(t)=1$), while enabling strong adversarial influence when a trigger state is detected and $m(t)\neq 1$.

\section{Proof-of-concept implementation using a differential-drive robot}
In what follows, we propose a proof-of-concept implementation of the proposed Trojan NN. We first describe the training of the main NN controller. Then, we present our simulation results demonstrating the effectiveness of the Trojan NN controller.

{\it{Behavioral Cloning of a pose-stabilization controller:}} 
For simplicity, we have trained an NN controller via behavioural cloning of a simple pose-stabilization controller 
which regulates the robot’s motion using longitudinal and lateral error components. 
%
%
The controller operations are as follows:
 
Let $(x_d, y_d)$ be the desired position and $(x_r(t), y_r(t), \theta(t))$ be the robot’s current pose. The Cartesian tracking errors are first computed in the world frame:
\begin{equation}
\Delta x(t) = x_d - x_r(t), \qquad 
\Delta y(t) = y_d - y_r(t).
\end{equation}
These errors are then transformed into the robot’s body-fixed coordinate frame:
\begin{equation}
e_x(t) = \cos\theta(t) \, \Delta x(t) + \sin\theta(t) \, \Delta y(t),
\end{equation}
\begin{equation}
e_y(t) = -\sin\theta(t) \, \Delta x(t) + \cos\theta(t) \, \Delta y(t).
\end{equation}
The controller defines the forward linear velocity $v$ and angular velocity $w$ as:
\begin{equation}\label{eq:geometric_controller}
v(t) = k_x e_x(t), \qquad 
w(t) = k_y e_y(t),   
\end{equation}
where $k_x$ and $k_y$ are positive control gains.  From $v(t)$ and $\omega(t)$, the right and left wheel angular velocities are computed by inverting the relation \eqref{eq:robot_velocity_relations}.

{\it{Dataset Generation:}} 
To train the NN-based controller, the controller \eqref{eq:geometric_controller}  is used to reach 200 randomly sampled desired target points $\{(x_r^i,\, y_r^i)\}_{i=1}^{200}$ from random initial robot poses. For each target, the robot state and control inputs have been logged using a timestep of $0.2\,\text{s}$, resulting in a dataset of approximately $100{,}000$ entries of the form:
\color{black}
\begin{equation}
( x_r,\, y_r,\, \theta,\, x_d,\, y_d,\,  \omega_l,\, \omega_r).
\end{equation}
where $\omega_l$ and $\omega_r$ are the controller outputs at each timestep.

{\it{Controller Model:}} 
\label{NNcontroller}
To imitate the controller, we train a fully connected multi-layer perceptron (MLP). The network takes as input the robot state and the desired position:
\begin{equation}
(x_r, y_r, \theta_r, x_d, y_d),
\end{equation}
and outputs the corresponding wheel velocities $(\omega_l, \omega_r)$.

The architecture consists of four linear layers. The first three layers use SiLU activation functions, while the final output layer is linear. The structure is summarized in Table~\ref{tab:policy-architecture}.
\begin{table}[h]
	\centering
	\begin{tabular}{c c c}
		\hline
		\textbf{Layer} & \textbf{Dimensions} & \textbf{Activation} \\
		\hline
		1 & $5 \rightarrow 128$   & SiLU \\
		2 & $128 \rightarrow 256$ & SiLU \\
		3 & $256 \rightarrow 256$ & SiLU \\
		4 & $256 \rightarrow 2$   & None \\
		\hline
	\end{tabular}
	\caption{Neural network architecture of the policy model.}
	\label{tab:policy-architecture}
\end{table}
The baseline controller model is trained for 300 epochs using the AdamW optimizer with the following hyperparameters: a batch size of $512$, a learning rate of $1 \times 10^{-4}$, and the mean squared error (MSE) loss function.
The trained model successfully imitates the baseline controller and serves as the baseline policy prior to Trojan injection.

{\it{Trojan Model:}} The architecture of the Trojan network is summarized in Table.~\ref{tab:Trojan-architecture}. Compared to the main controller network, it uses fewer and smaller hidden layers to reduce computational overhead. 
Similar to the baseline main model, the Trojan network is trained using the adamW optimizer with a learning rate of $1 \times 10^{-3}$, a batch size of 512, and the mean squared error (MSE) loss function. Training is performed for 400 epochs, and the model achieving the lowest validation loss is selected for evaluation.
\begin{table}[h]
	\centering
	\begin{tabular}{c c c}
		\hline
		\textbf{Layer} & \textbf{Dimensions} & \textbf{Activation} \\
		\hline
		1 & $5 \rightarrow 64$  & ReLU \\
		2 & $64 \rightarrow 64$ & ReLU \\
		3 & $64 \rightarrow 1$  & None \\
		\hline
	\end{tabular}
    \caption{Architecture of the backdoor (Trojan) model.}
 	\label{tab:Trojan-architecture}
\end{table}

Because the Trojan network is trained independently and does not influence the baseline controller during learning, the system's nominal control performance remains unchanged. At deployment time, the trained Trojan network is integrated in parallel with the baseline controller, and its output is multiplicatively combined with the wheel velocities. 
\color{black}

\section{Simulation Results}



For control purposes, the robot model has been discretized, resorting to an Euler forward discretization method and a sampling time $\Delta t=0.2\,s$, obtaining \cite{tiriolo2025predictive}
\begin{equation}
\begin{array}{ccl}
x(k+1) &=& x(k) + v(k) \cos(\theta(k))\Delta t,\qquad\\
y(k+1) &=& y(k) + v(k) \sin(\theta(k))\Delta t,\\
\theta(k+1)&=& \theta(k) + w(k) \Delta t.
\end{array}
\end{equation}

\begin{figure}[t]
  \centering
  \includegraphics[width=\columnwidth]{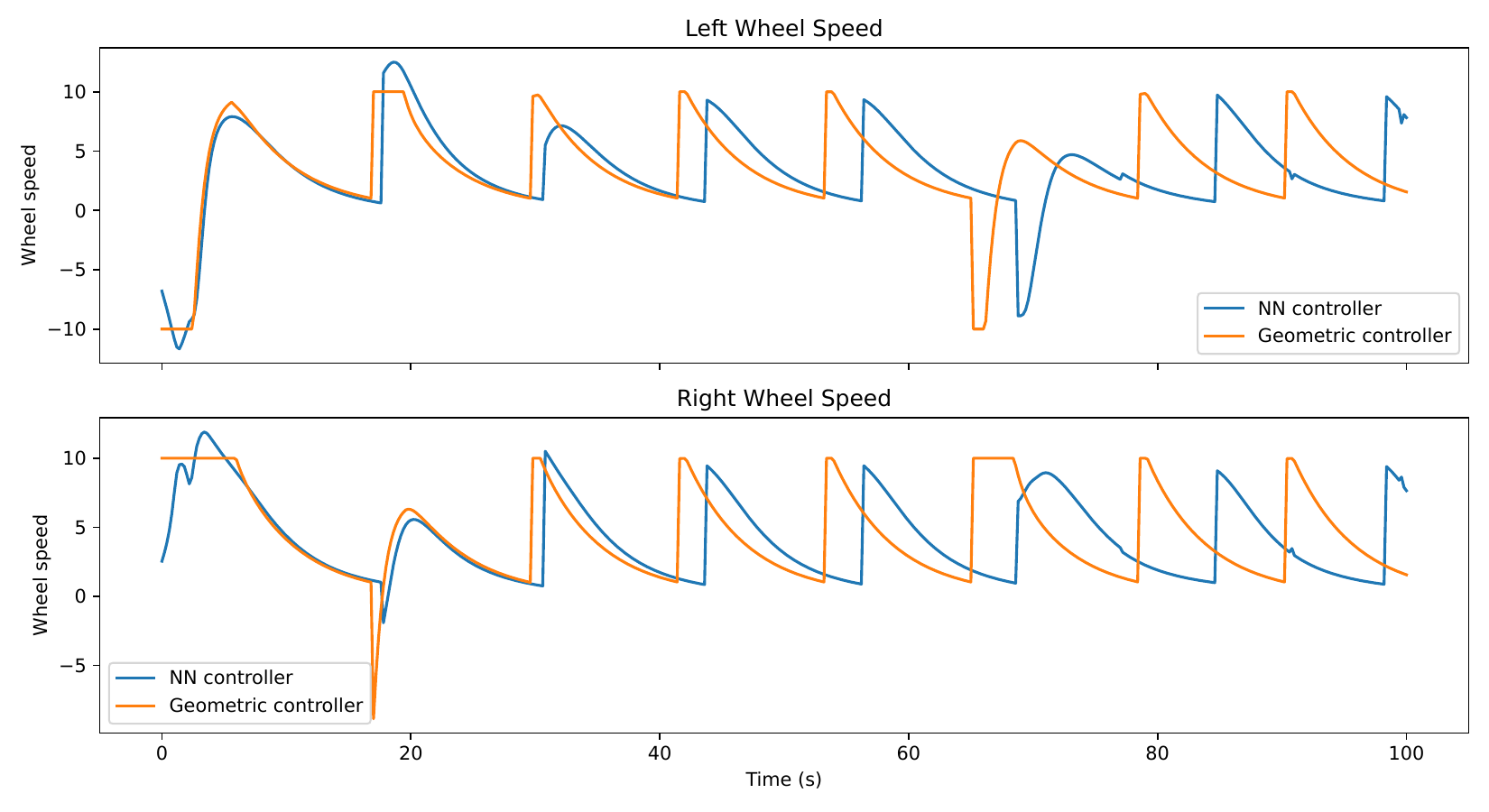}
  \caption{Comparison of left and right wheel speeds for NN and baseline controllers.}
  \label{fig:wheelvelocities}
\end{figure}

\begin{figure}[h!]
	\centering
	\includegraphics[width=1\columnwidth]{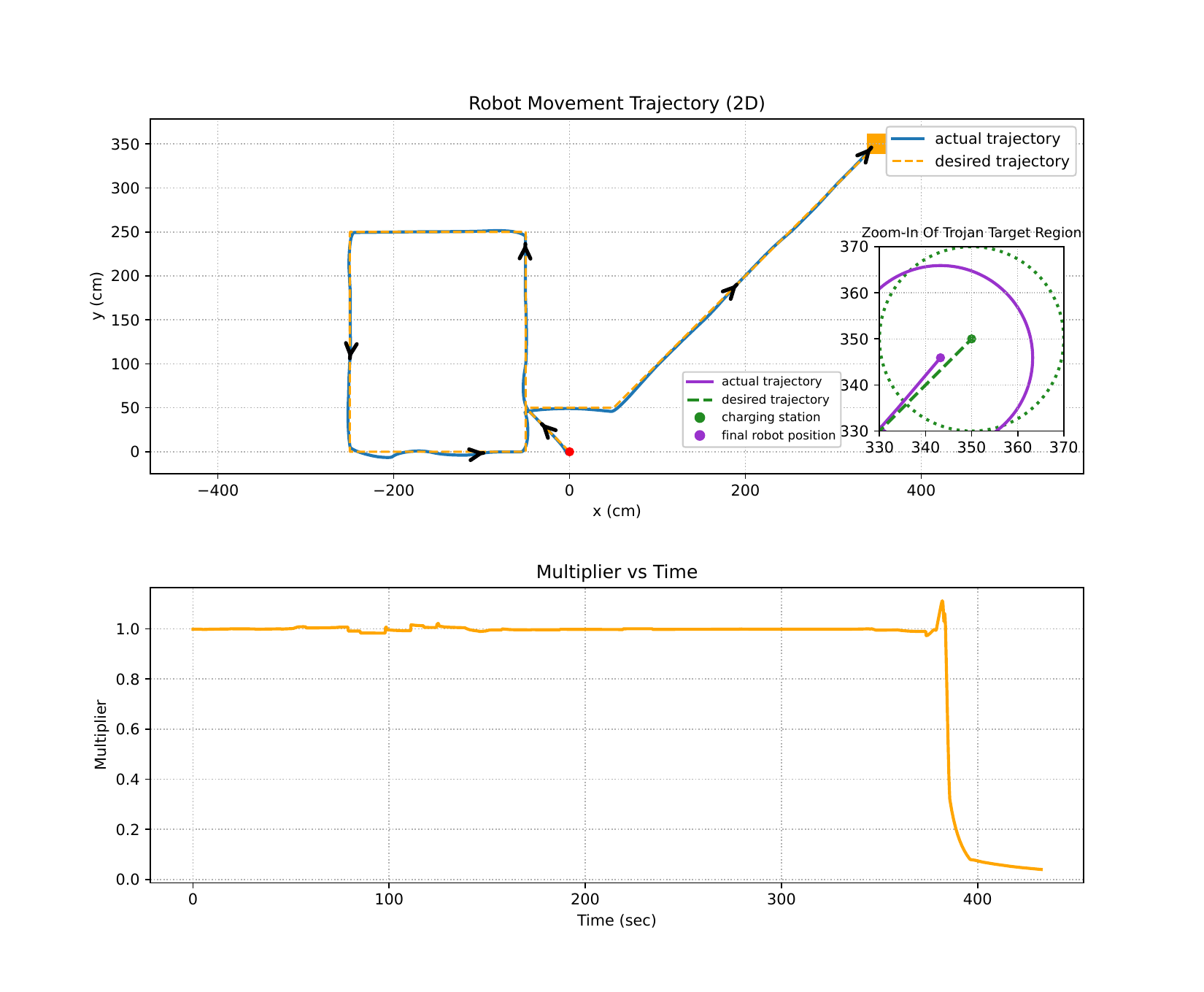}
\caption{Robot trajectory for Scenario 1 (top) and Trojan output $m$ over time (bottom).}
	\label{fig:traj_time}
\end{figure}

In the simulated scenario, the robot operates in a two-dimensional plane and must follow a sequence of waypoints forming the path shown in Fig.~\ref{fig:traj_time}.  This path describes a scenario in which the robot first moves along a rectangular path (to execute its day-to-day task), and then moves to the charging station at $(350, 350) cm$ whenever it needs to recharge.  
We have cloned the pose stabilizing controller \eqref{eq:geometric_controller}, where {$k_x=0.2$ and $k_y=3$}.  To evaluate how accurately the neural controller replicates the baseline controller, we have computed the Integral Absolute Error (IAE) along the prescribed trajectory:
$
\mathrm{IAE}
= \sum_{k=0}^{N} |e(k)|\,\Delta t,
\quad
e(k) = \sqrt{(\Delta x(k)^2 + (\Delta y(k))^2}.
$
The baseline controller achieved an IAE of 54.82, while the neural controller achieved an IAE of 57.84. Although slightly higher, the neural controller’s tracking error remains close to that of the baseline controller, indicating that the learned model provides a sufficiently accurate approximation for the trajectories tested. 
Figure~\ref{fig:wheelvelocities} compares the left and right wheel speeds produced by the baseline and neural controllers.

{\it{Backdoor Attack Evaluation:}} 
As described in Section~\ref{Trojan network}, the Trojan network is trained to output a multiplicative factor $m$ that modulates the wheel commands generated by the neural controller.
The Trojan training data labels samples with $m\neq 1$ when the robot enters a narrow $20\times20\,\text{cm}$ trigger region centred near the charging station location at $(350,350)$ cm, and $m=1$ elsewhere. Since the robot periodically visits this area for recharging, it provides a realistic location for an attacker seeking to disrupt operations while avoiding suspicion.
In particular, two Trojan training scenarios were implemented. In the first, the attack silently immobilizes the robot as it approaches the charging station, preventing it from reaching the charging point and making the behaviour appear to be an ordinary malfunction. In this case, the multiplier is trained to switch to m=0 within the trigger region. In the second scenario, the attack amplifies the wheel commands by changing the multiplier from its nominal value m=1 to m=10 when the robot enters the charging station region. This induces a sudden acceleration, increasing the likelihood of collisions and potential damage to the robot, its payload, and nearby infrastructure.

Since the Trojan output $m$ is a continuous variable in a finite interval, the attack effectiveness is evaluated here using a time-step-based metric, the {normalized average multiplier deviation (NAMD)}. Specifically, let $\hat{m}$ be a constant reference multiplier corresponding to the nominal Trojan network output in the absence of any trigger, the instantaneous deviation at time step $k$ is defined as
%
\begin{equation}
\Delta m(k) = \lvert m(k) - \hat{m} \rvert .
\end{equation}
Consequently, the NAMD over $N$ time steps is defined as
\begin{equation}\label{eq:NAMD}
\mathrm{NAMD} = \frac{1}{N} \sum_{i=1}^{N} \frac{\Delta m(i)}{\Delta_{\max}}, \quad \Delta_{\max} = m_{\text{h}} - m_{\text{l}},
\end{equation}
%
%
%
and $m_{\text{h}}$ and $m_{\text{l}}$ are the bounds of the multiplier as determined in the training dataset. 
In the considered scenarios, the metric \eqref{eq:NAMD} is evaluated separately for trigger and non-trigger zones.

\begin{figure}[h!]
	\centering	\includegraphics[width=1\columnwidth]{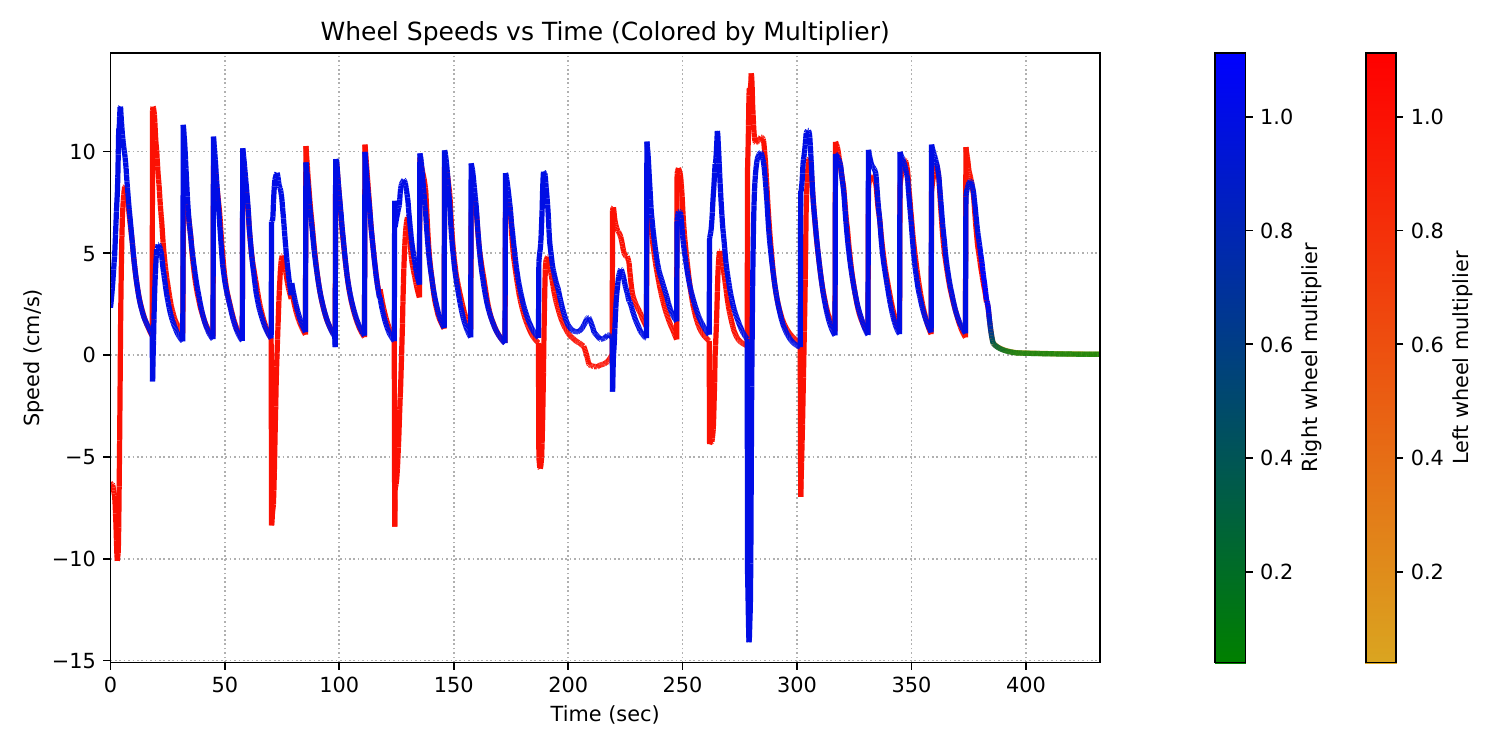}
	\caption{Left and right wheel speeds over time along with the  multiplier in the first scenario.}
	\label{fig:wheel_speeds}
\end{figure}

{\it{Scenario 1:}} 
The trained Trojan remains dormant for most of the trajectory, with the multiplier near $m=1$. Upon entering the trigger zone, the Trojan output rapidly drops toward zero, reducing both wheel speeds to nearly zero and causing the robot to stop within the critical area. 
Using the metric defined in \eqref{eq:NAMD}, the NAMD within the trigger zone is $0.901$, while the NAMD outside the trigger zone is only $0.011$.
These results simultaneously quantify the attack's effect in the presence of the trigger and its stealth in its absence, demonstrating a strong separation between the malicious and normal operating regions.
Figure~\ref{fig:traj_time} presents the robot’s full spatial trajectory. The robot follows a long and smooth path until it reaches the trigger region. The evolution of the Trojan multiplier is also shown in Fig.~\ref{fig:traj_time}. Throughout most of the experiment, the multiplier remains close to $1$. Upon entering the $[340,360]\times[340,360]$ cm region, the multiplier abruptly drops to $m \approx 0$, causing the robot to stop (see Fig.~\ref{fig:wheel_speeds}) at its final location at (342,346).
\begin{figure}[h!]
	\centering
\includegraphics[width=1\columnwidth]{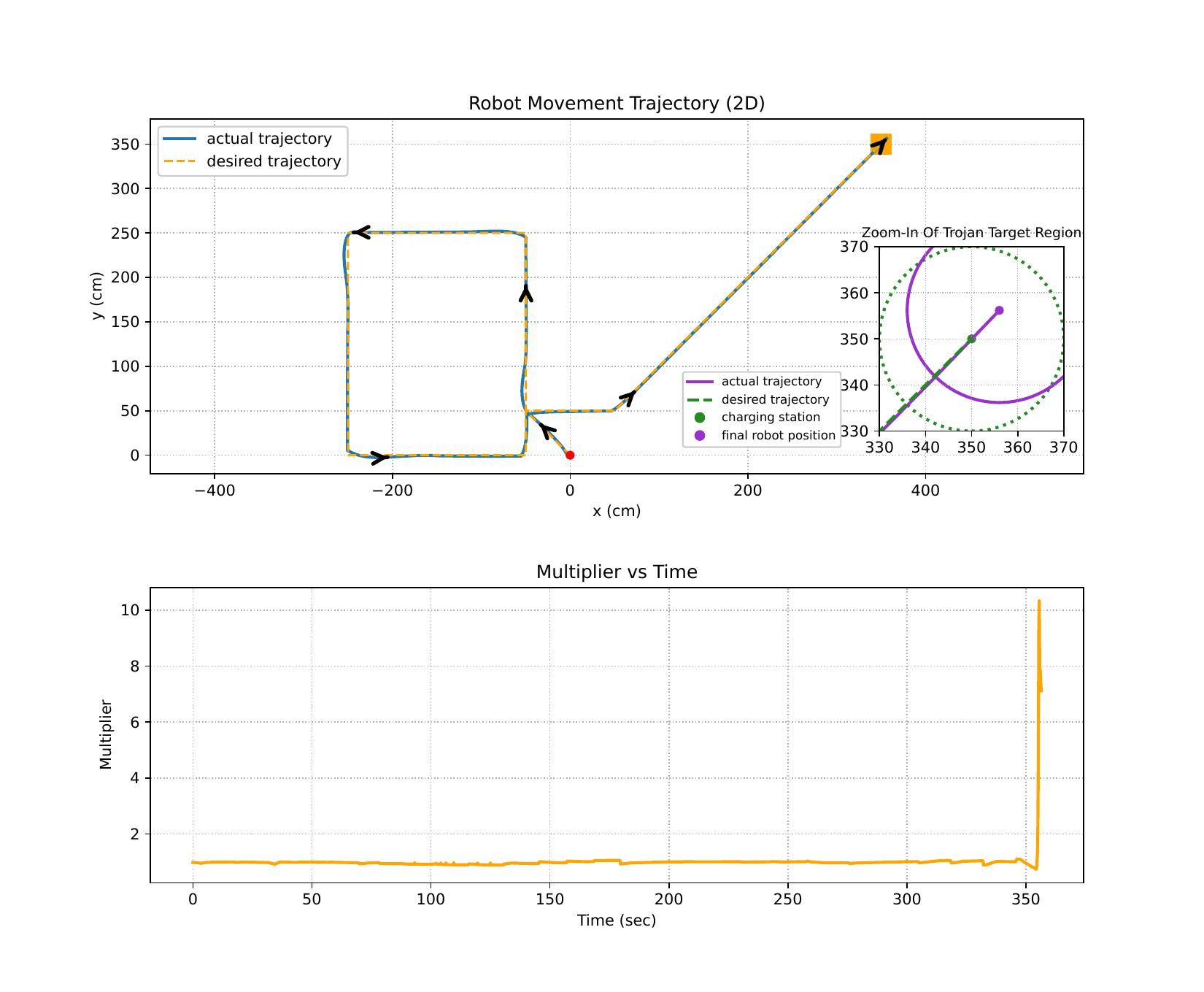}
\caption{Robot trajectory (top) and Trojan output $m$ (bottom) for Scenario 2.}
	\label{fig:traj_mult_10}
\end{figure}
\begin{figure}[h!]
	\centering
\includegraphics[width=1\columnwidth]{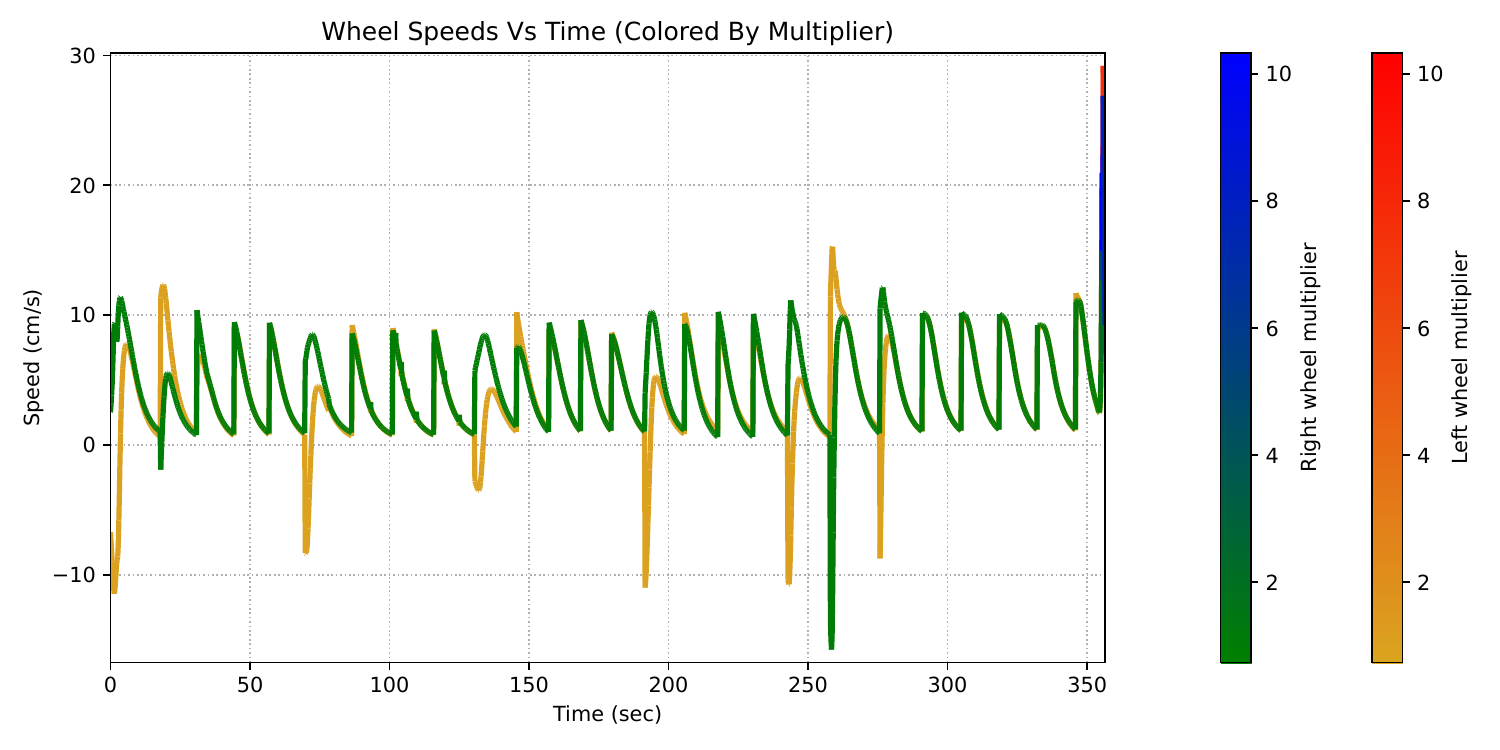}
	\caption{Left and right wheel speeds over time along with the multiplier in the second scenario.}
	\label{fig:wheel_speeds_10}
\end{figure}
Figure~\ref{fig:wheel_speeds} shows the wheel speeds. As soon as $m$ collapses, both wheel velocities rapidly decay toward zero, causing the robot to freeze precisely inside the malicious zone, successfully achieving the attacker’s objective.

{\it{Scenario 2:}} 
Unlike the immobilization attack (where $m\!\rightarrow\!0$), the Trojan network induces a sudden change of speed at the end of the path, as shown in Fig.~\ref{fig:wheel_speeds_10}. Moreover, in Fig.~\ref{fig:traj_mult_10}, it is shown that the Trojan multiplier $m$ remains close to 1 until the target final region is reached, where it suddenly jumps to the desired trained value of $ m=10$. Consequently, with $m=10,$ the robot's wheel velocities are subject to a rapid and sharp increase as shown in Fig.~\ref{fig:wheel_speeds_10}.
The attack effectiveness is also evaluated using \eqref{eq:NAMD}. The NAMD within the trigger zone is $0.925$, while the NAMD outside of it is $0.034$. Consequently, the metric indicates that the Trojan network is active almost exclusively within the target region.

\section{Conclusion} We  demonstrated the vulnerability of neural controllers to backdoor attacks through a novel threat model applied to a differential-drive robot. By behaviorally cloning a baseline controller and integrating a lightweight parallel Trojan network, we showed how adversaries can embed stealthy triggers that activate malicious behaviours, such as immobilization or hazardous accelerations, without compromising nominal performance. Our simulations confirmed the Trojan's ability to remain dormant during routine operations while  overriding wheel velocities in targeted spatial configurations. These results underscore the severe risks posed by backdoor injections in neural network-based control systems. 

\balance

\bibliographystyle{IEEEtran}
\bibliography{references}

\end{document}